# Climate predictions: the chaos and complexity in climate models

D.T. Mihailović[1,*], G. Mimić[2] and I. Arsenić[1]
[1]Faculty of Agriculture, University of Novi Sad, Dositej Obradovic Sq. 8, 21000 Novi Sad, Serbia
[2] Department of Physics, Faculty of Sciences, University of Novi Sad, Dositej Obradovic Sq. 5, 21000 Novi Sad, Serbia

Abstract. Some issues which are relevant for the recent state in climate modeling have been considered. A detailed overview of literature related to this subject is given. The concept in modeling of climate, as a complex system, seen through the Gödel's Theorem and Rosen's definition of complexity and predictability is discussed. It is pointed out to occurrence of chaos in computing the environmental interface temperature from the energy balance equation given in a difference form. A coupled system of equations, often used in climate models is analyzed. It is shown that the Lyapunov exponent mostly has positive values allowing presence of chaos in this system. The horizontal energy exchange between environmental interfaces, which is described by the dynamics of driven coupled oscillators, is analyzed. Their behavior and synchronization, when a perturbation is introduced in the system, as a function of the coupling parameter, the logistic parameter and the parameter of exchange, was studied calculating the Lyapunov exponent under simulations with the closed contour of $N = 100$ environmental interfaces. Finally, we have explored possible differences in complexities of two global and two regional climate models using their output time series by applying the algorithm for calculating the Kolmogorov complexity.

## 1 Introduction

Among the most interesting and fascinating phenomena that are predicted/predictable are the chaotic ocean/atmosphere/land system called weather and its long time average - climate. While weather is not predictable beyond a few days, aspects of the climate may be predictable for years, decades, and perhaps longer [1]. These two phrases clearly summarise the current opinion and state in climate modeling community that deals with the aforementioned subjects. However, the question of the weather and climate modeling and predictability has been initiated in early sixties of the 20[th] century, which was elaborated in pioneering works by Edward N. Lorenz [2-5]. He was the first person in the scientific world who explicitly pointed out the following points related to the nonlinear dynamics in atmospheric motion: (i) question of prediction and predictability, (ii) importance of understanding the nonlinearity in modeling procedure, (iii) demand for discovery of chaos and (iv) careful consideration of sensitivity of differential equations in modeling system on initial conditions. Subsequent three decades after appearance of these papers, have been characterised by strong interest for predictability of weather and climate on theoretical and practical level. The following topics have been set in the focus: (1) dynamics of error growth; (2) linear and nonlinear systems (normal modes and optimal modes, nonlinear geophysical systems and scale selection in error growth); (3) predictability of systems with many scales; (4) limit of predictability; (5) weather predictability (growth of errors in General Circulation Models (GCMs) based on Lorenz's analysis); (6) predictability from analogs (targeted observations); (7) climate predictability (predictability of time-mean quantities, predictability of the second kind) and potential predictability; (8) seasonal mean predictability and (9) El Niño-Southern Oscillation (ENSO) chaos, predictability of coupled models and decadal modulation of predictability [6-18]. Because the focus of our paper is complexity and predictability in climate modeling, we finish this overview with the comment by Orell (2003): "Prediction problems have been described by Lorenz as falling into two categories. Problems that depend on the initial condition, such as short- to medium-range weather forecasting, are described as predictions of the first kind, while problems that depend on boundary rather than initial conditions, such as, in many cases, the longer-term climatology, are referred to as predictions of the second kind. Both kinds of prediction will be affected by error in the model equations used to approximate the true system" [19-21].

---

[*] Corresponding author: D. T. Mihailović
 Email address: guto@polj.uns.ac.rs

Earth's atmosphere has evolved into a complex system in which life and climate are intricately interwoven. The interface between Earth and atmosphere as a "pulsating biophysical organism" is a complex system itself. The term *complex system* we use in Rosen's sense (Rosen, 1991) as it was explicated in the comment by Colier (2003): "In Rosen's sense a complex system cannot be decomposed non-trivially into a set of part for which it is the logical sum. Rosen's modeling relation requires this. Other notions of modeling would allow complete models of Rosen style complex systems, but the models would have to be what Rosen calls *analytic*, that is, they would have to be a logical product. Autonomous systems must be complex. Other types of systems may be complex, and some may go in and out of complex phases" [22, 23]. Also, we will explain in which sense the term *complexity* will be used in further text. Usually, that is an ambiguous term, sometimes used [22] to refer to systems that cannot be modeled precisely in all respects. However, following Arshinov and Fuchs (2003) the term "complexity" has three levels of meaning [24]: (1) there is self-organization and emergence in complex systems [25], (2) complex systems are not organized centrally, but in a distributed manner; there are many connections between the system's parts [25, 26], (3) it is difficult to model complex systems and to predict their behaviour even if one knows to a large extent the parts of such systems and the connections between the parts [25, 27]. The complexity of a system depends on the number of its elements and connections between the elements (the system's structure). According to this assumption, Kauffman (1993) defines complexity as the "number of conflicting constraints" in a system [26], Heylighen (1996) says that complexity can be characterized by a lack of symmetry (symmetry breaking) which means that "no part or aspect of a complex entity can provide sufficient information to actually or statistically predict the properties of the others parts" [28] and Edmonds (1996) defines complexity as "that property of a language expression which makes it difficult to formulate its overall behavior, even when given almost complete information about its atomic components and their inter-relations" [29]. Aspects of complexity are things, people, number of elements, number of relations, non-linearity, broken symmetry, non-holonic constraints, hierarchy and emergence [30].

Generally, predictability refers to the degree that a correct forecast of a system's state can be made either qualitatively or quantitatively. For example, while the second law of thermodynamics can tell us about the equilibrium that a system will evolve to, and steady states in dissipative systems can sometimes be predicted, there exists no general rule to predict the time evolution of systems far from equilibrium, i. e. chaotic systems, if they do not approach some kind of equilibrium. Their predictability usually deteriorates with time. To quantify predictability, the rate of divergence of system trajectories in phase space can be measured (Kolmogorov-Sinai entropy, Lyapunov exponents).

Lorenz (1984) discussed several issues in the predictability of weather systems [31]. According to him predictability is defined as the degree of accuracy with which it is possible to predict the state of weather system in the near and also the distant future (predictability in Lorenz's sense). In this paper it is assumed that weather predictions are made on the basis of imperfect knowledge of a weather system's present and past states. This rather general statement is comprehensively elaborated by Hunt (1999) [20]. He described the fundamental assumptions and current methodologies of the two main kinds of environmental forecast (i.e., weather forecast); the first is valid for a limited period of time into the future and over a limited space–time "target", and is largely determined by the initial and preceding state of the environment, such as the weather or pollution levels, up to the time when the forecast is issued and by its state at the edges of the region being considered; the second kind provides statistical information over long periods of time and/or over large space–time targets, so that they only depend on the statistical averages of the initial and "edge" conditions. Environmental forecasts depend on the various ways that models are constructed. These range from those based on the "reductionist" methodology (i.e., the combination of separate, scientifically based, models for the relevant processes) to those based on statistical methodologies, using a mixture of data and scientifically based empirical modeling. For example, limitations of the predictability in the world of atmospheric motions are concisely discussed in paper by James (2002) [32]. In this paper it is numerically considered the predictability of a forced nonlinear system, proposed by Lorenz, as a compelling heuristic model of the mid-latitude global circulation.

The above insight of the predictability underlined in the context of the "environmental predictability" (primarily linked to the climate change issues), we finish with the question: *Can we significantly "improve" the weather/climate predictions comparing to the level they currently reached*? The answer can not be strictly elaborated with either *yes* or *no*. An optimistic and acceptable attitude, that prefers option *yes*, is concisely written down by Hunt (1999) as the phrase: "We concluded that philosophical studies of how scientific models develop and of the concept of determinism in science are helpful in considering these complex issues" [20]. If we give advantage to the option *no* then we do not close the door for the first option. It only means that there exists limitation of the modeling attempts on an epistemological level. To show that, we will use the Gödel's Incompleteness Theorem about Number Theory [33]. Basically it says that no matter how one tries to formalize a particular part of mathematics, syntactic truth in the formalization does not coincide with the set of truths about numbers. In other word Gödel's Theorem shows that formalizations are part of mathematics, but not all of mathematics. There are many ways to look and "read" Gödel's Theorem. One exclusive way is offered by Rosen (1985) [34]. According to him the first thing to bear in mind is that both Number Theory and any formalization of it are both systems of entailment. It is the *relation* between them, or more specifically, the extent to which these schemes of entailment can be brought into congruence, that is of primary interest. The establishment of such congruencies, through the positing of referents in one of them for elements of the other, is the essence of the *modeling relation*. In a precise sense, this theorem asserts that a formalization which all entailment is syntactic entailment is too *impoverished in entailment* to be congruent to Number Theory, no matter how we try to establish such congruence. This kind of situation is termed *complexity* by Rosen (1977) [35]. Namely, in this light, Gödel's Theorem says that Number Theory is more *complex* than any of its formalization, or equivalently, that formalizations, governed by syntactic inference alone, are *simpler* than Number Theory. To reach Number Theory from its formalizations, or more generally, to reach a complex system from simpler one, requires some kind of limiting processes.

Let us return to the question we were asking ourselves after we had shortly considered climate modeling (i.e., predictability) beyond the complexity. To our mind there is a significant space for "improvement" of models and their capabilities to provide good forecasts. It can be done only if the modeling attempts are directed towards the following steps: from structures and states to processes and functions; from self-correcting to self-organizing systems; from hierarchical steering to participation, from conditions of equilibrium to dynamic balances of non equilibrium; from single trajectories to bundles of trajectories; from linear causality to circular causality; from predictability to relative chance; from order and stability to instability, chaos and dynamics; from certainty and determination to a larger degree of risk, ambiguity and uncertainty; from reductionism to emergetism and from being to becoming.

In this paper we address three issues that, to our mind, are important for further improvements in designing the climate models. (1) The phenomenon of chaos in computing the environmental interface temperature from the energy balance equation which will be considered through the question how to replace given differential equations by appropriate difference equations in climate simulations? (Section 2). (2) The synchronization of energy exchange between environmental interfaces in dependence on perturbation of environmental parameters (Section 3) and (3) complexity analysis of the climate model output time series which is elaborated in Section 4. In Section 5 we give concluding remarks.

**2. Energy balance equation: Occurence of chaos in computing the environmental interface temperature**

*2.1 Background*

Traditional mathematical analysis of physical systems tacitly assumes that integers and all real numbers, no matter how large or how small, are physically possible and all mathematically possible trajectories are physically possible [36]. Traditionally, this approach has worked well in physics and in engeneering but it does not lead to a very good understanding of chaotic systems, which, as is now known, are extremely important in the study of real world-phenomena ranging from weather to biological systems. In this paper we deal with one issue in modeling pathways in meteorology as well as in physics, biology and chemistry, i.e. environmental sciences in their broadest context [37], in particular in autonomous dynamical systems, which are common subject under consideration in climate modeling. Namely, we consider how to replace given differential equations by appropriate difference equations in environmental modelling and thus in climate simulations [38].

According to van der Vaart many models for environmental problems have been and will be built in the form of differential equations or systems of such equations [38]. With the advent of computers one has been able to find (approximate) solutions for equations that used to be intractable. Many of the mathematical techniques have been used in this area to replace given differential equations by appropriate difference equations. So a huge effort has been invested into choice of appropriate difference equations whose solutions are "good" approximations to the solutions of the given differential equations. This question includes a requirement for better understanding of the fundamental problem: interrelations between classical continuum mathematics and reality in different sciences. For many atmospheric phenomena the "continuum" type of thinking, that is at the basis of any differential equation, is not natural to the phenomenon, but rather constitutes an approximation to a basically discrete situation: in much work of this type the "infinitesimal step lengths" handled in the reasoning which lead us to the differential equation, are not really thought of as infinitesimally small, but as finite; yet, in the last stage of such reasoning, where the differential equation rises from the differentials, these "infinitesimal" step lengths go to zero: that is where above-mentioned approximation comes in. Under this kind of circumstances, it seems more natural *to build the model* as a discrete difference equation from the start, without going through the painful, doubly approximative process of first, during the modeling stage, finding a differential equation to approximate a basically discrete situation, and then, for numerical computing purposes, approximating that differential equation by a difference scheme [36].

In this section we analyze the energy balance equation in procedure of computing the environmental interface temperature and the deeper soil layer temperature commonly used in climate models. The environmental interface is defined as *interface between two biotic or abiotic environments that are in relative motion and exchange energy, matter and information through physical, biological and chemical processes, fluctuating temporally and spatially regardless of space and time scale* [39]. There are a lot of examples of environmental interfaces in the nature, but here we deal with, the ground surface, where there exist all three mechanisms of energy transfer; incoming and outgoing radiation, convection of heat and moisture into the atmosphere and conduction of heat into deeper soil layers of ground (Figure 1) [40]. Parameterization of these processes is of great importance for environmental models of different spatial and temporal scales, and thus climate ones. In the paper by Mihailović and Mimić (2012) it is shown that ground surface is treated as a complex system in which chaotic fluctuations occur while we compute its temperature [41]. This system, as an actual dynamic system, is very sensitive to initial conditions and arbitrarily small perturbation of the current trajectory that may lead to its unpredictable behavior. In the aforementioned paper the lower boundary condition, i.e. the deeper soil layer temperature was constant, but it can also vary in time making with the energy balance equation a coupled system of equations. That system, often used in environmental models, is of interest to be analyzed by the methods of nonlinear dynamics.

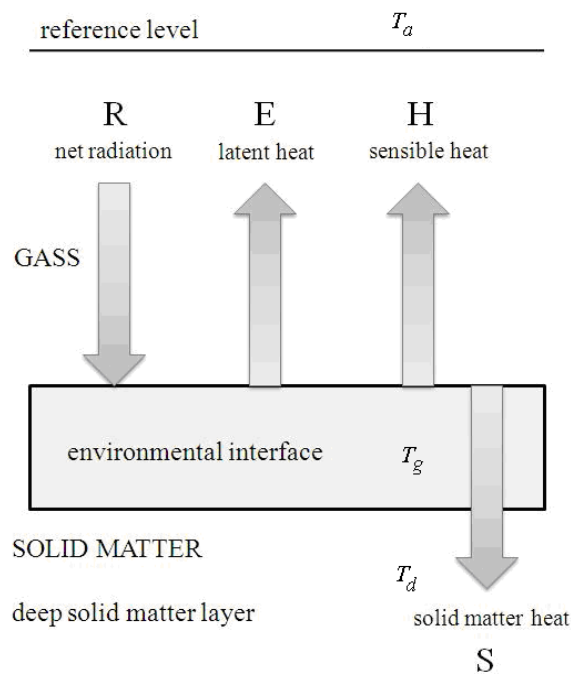

Figure 1. Terms in energy balance equation.

Having in mind those facts, in this section we: (i) perform a nonlinear dynamical analysis of coupled system for computing the environmental interface temperature and the deeper soil layer temperature and (ii) examine behavior of the coupled system in dependence on the main system parameters, in order to show the possible occurence of the chaos in computing the environmental interface temperature. Firstly, we consider difference form of the energy balance equation and deeper soil layer temperature equation transforming them into the coupled system with the corresponding parameters and then we analyze behavior of the solutions of the coupled system and we have examined domains of stability using the Lyapunov exponent.

*2.2 Physical background and derivation of the coupled system*

One of the most important conditions for functioning of any complex system is a proper supply of the system with energy. Dynamics of energy flow is based on the energy balance equation [40]. As we mentioned before, environmental interface is a complex system. General difference form of energy balance equation for the ground surface as an environmental interface is

$$C_g \frac{\Delta T_g}{\Delta t} = R_{net} - H - \lambda E - G \qquad (1)$$

where $T_g$ is the ground surface temperature, $\Delta t$ is the time step, $C_g$ is the soil heat capacity, $R_{net}$ is the net radiation, $H$ is the sensible heat flux, $\lambda E$ is the latent heat flux and $G$ is the heat flux into the ground. First, we assume that the net radiation is given as in [42], i.e.

$$R_{net} = C_R (T_g - T_a) \qquad (2)$$

where $T_a$ is the air temperature at some reference level and $C_R$ is the coefficient for the net radiation term. Second, we make expansion of the exponential term in the expression for latent heat flux

$$\lambda E = C_L d \left[ b(T_g - T_a) + \frac{b^2}{2}(T_g - T_a)^2 \right], \qquad (3)$$

where $C_L$ is the water vapour transfer coefficient, $b = 0.06337 \,^{\circ}C^{-1}$, $d$ is parameter which occurs in expanding the series [43]. Further, the conduction of the heat into the soil can be written in the form

$$G = C_D (T_g - T_d), \qquad (4)$$

where $C_D$ is the coefficient of the heat conduction while $T_d$ is the temperature of the deeper soil layer. The sensible heat flux $H$ can be parameterized as

$$H = C_H (T_g - T_a), \qquad (5)$$

where $C_H$ is the sensible heat transfer coefficient. The prognostic equation for temperature of the deeper soil layer $T_d$ is

$$\frac{\Delta T_d}{\Delta t} = \frac{1}{\tau}(T_g - T_d), \qquad (6)$$

where $\tau = 86400\ s$. After collecting all terms (2)-(6), the coupled system takes the form

$$C_g \frac{\Delta T_g}{\Delta t} = C_R(T_g - T_a) - C_H(T_g - T_a) - C_L d[b(T_g - T_a) \\ + \frac{b^2}{2}(T_g - T_a)^2] - C_D(T_g - T_d) \tag{7}$$

$$\frac{\Delta T_d}{\Delta t} = \frac{1}{\tau}(T_g - T_d) \ . \tag{8}$$

More details about the nature and the range of physical parameters $C_R$, $C_L$, $C_D$ and $C_H$ can be found in [44]. Now, using the time scheme forward in time (n indicates the time step) and dividing both sides of Eqs. (7) and (8) with the constant temperature $T_0$ (for example, value of mean Earth temperature, i.e. $T_0 = 288K$) we get

$$\frac{T_g^{n+1} - T_a^n}{T_0} = \frac{T_g^n - T_a^n}{T_0} + \frac{\Delta t}{C_g} C_R \frac{T_g^n - T_a^n}{T_0} - \frac{\Delta t}{C_g} C_H \frac{T_g^n - T_a^n}{T_0} - \frac{\Delta t}{C_g} C_L bd \frac{T_g^n - T_a^n}{T_0} \\ - \frac{\Delta t}{C_g} C_L dT_0 \frac{b^2}{2} \frac{(T_g^n - T_a^n)^2}{T_0^2} - \frac{\Delta t}{C_g} C_D \frac{T_g^n - T_a^n}{T_0} + \frac{\Delta t}{C_g} C_D \frac{T_d^n - T_a^n}{T_0} \tag{9}$$

$$\frac{T_d^{n+1} - T_a^n}{T_0} = \frac{T_d^n - T_a^n}{T_0} + \frac{\Delta t}{\tau} \frac{T_g^n - T_a^n}{T_0} - \frac{\Delta t}{\tau} \frac{T_d^n - T_a^n}{T_0} \ . \tag{10}$$

Finally, introducing replacements $x = (T_g - T_a)/T_0$ and $y = (T_d - T_a)/T_0$, where $x$ is the dimensionless environmental interface temperature and $y$ is the dimensionless deeper soil layer temperature, we reach the coupled system

$$x_{n+1} = Ax_n - Bx_n^2 + Cy_n \tag{11}$$

$$y_{n+1} = Dx_n + (1-D)y_n, \tag{12}$$

where $A = 1 + \frac{\Delta t}{C_g}(C_R - C_H - C_L bd - C_D)$, $B = C_L dT_0 \frac{b^2 \Delta t}{2C_g}$, $C = \Delta t \frac{C_D}{C_g}$ and $D = \frac{\Delta t}{\tau}$. Introducing the replacement $x_{1,n} = Ax_n / B$, where $x_1$ is the modified dimensionless environmental interface temperature and $x_{2,n} = y_n$, we can write

$$x_{1,n+1} = Ax_{1,n}(1 - x_{1,n}) + \frac{CB}{A} x_{2,n} \tag{13}$$

$$x_{2,n+1} = \frac{DA}{B} x_{1,n} + (1-D)x_{2,n} . \tag{14}$$

Analysis of values of parameters $A$, $B$, $C$ and $D$, based on a large number of energy flux outputs from the land surface scheme runs, indicates that their values are ranged in the following intervals: (i) $A \in [0,4]$ and (ii) $B$, $C$ and $D$ are ranged in the interval $[0,1]$. Thus, $A$ is the logistic parameter, which from now will be denoted with $r$. All other groups of parameters in the system (13)-(14) have the values in the same interval $[0,1]$. Let us underline that under some circumstances those parameters can be equal. Correspondingly, we replaced all of them by introducing the coupling parameter $c$.

Finally, system (13)-(14) can be written in the form of coupled maps, i.e.,

$$x_{1,n+1} = rx_{1,n}(1 - x_{1,n}) + cx_{2,n} \qquad (15)$$

$$x_{2,n+1} = c(x_{1,n} + x_{2,n}) \qquad . \qquad (16)$$

Now we analyse the stability of physical solutions of coupled maps (15)-(16), using the Lyapunov exponent, which is a measure of convergence or divergence of near trajectories in phase space. Sign of Lyapunov exponent is characteristic of attractor type and for stable fixed point is negative, although for chaotic attractor is positive. Calculating Lyapunov exponent for the coupled system (15)-(16) with values of parameters $c \in (0.05, 0.1)$ and $r \in (3.6, 3.8)$, because we thought that will be interesting to investigate behavior of the system for small values of coupling parameter and high values of logistic parameter, we got results depicted in Figure 2 [45]. It is shown that the Lyapunov exponent mostly has positive values approving presence of chaos in this system, but there are still some strait regions where the Lyapunov exponent is negative and where the solutions of the coupled system are stable, i.e. domains of stability.

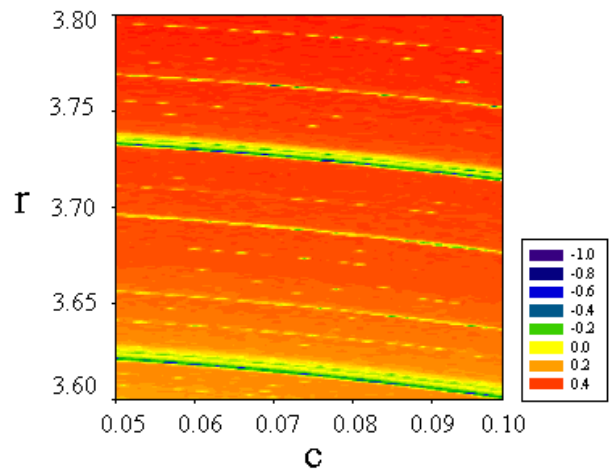

Figure 2. Lyapunov exponent of the coupled system (15)-(16), which shows presence of strait regions of stability in highly developed chaos.

## 3. Horizontal energy exchange between environmental interfaces

*3.1 Background*

There are three major sets of processes that must be considered when constructing a climate model: (i) radiative (the transfer of radiation through the climate system, e.g. absorption and reflection); (ii) dynamic (the horizontal and vertical transfer of energy, e.g. advection, convection and diffusion) and (iii) surface process (inclusion of processes involving land/ocean/ice, and the effects of albedo, emissivity and surface-atmosphere energy exchanges). If the nonlinearities in these processes are treated improperly then in designing the model, the complexity, and thus its reliability, will not be retained in the highest degree. In Section 2 we have considered surface-atmosphere energy exchanges with cadence on the phenomenon of a possible occurrence of the chaos in solving the energy balance equation for calculating the environmental interface temperature in climate models. Here, relying on Section 2 and using paper by Mihailović *et al.* (2012) we analyze the horizontal energy exchange between environmental interfaces which is described by the dynamics of driven coupled oscillators [46]. In order to study their behavior, when a perturbation is introduced in the system, as a function of the coupling parameter, the logistic parameter and the horizontal energy exchange intensity (parameter of exchange, in further text), we considered dynamics of two maps serving the diffusive coupling [46].

As noted above, the horizontal exchange of energy between environmental interfaces is considered as diffusion-like process. The dynamics of energy exchange behavior on environmental interface are typically expressed as a logistic map $\Phi(x) = rx(1-x)$, where $x$ is the dimensionless temperature of environmental interface and $r$ is a logistic parameter [45, 46]. However, we use an alternative form of this map, which includes a parameter $p$ that represents the horizontal energy exchange intensity (Figure 3). By introducing this parameter we formalize an intrinsic property of the environmental interfaces, which depends on the nature of the interface. The environmental interface dynamics are expressed here as a difference equation, so we avoid the double approximation of (i) finding a differential equation to approximate an essentially discrete process (during the modeling stage) and then (ii) approximating that differential equation by a difference scheme for numerical computing purposes [36, 38], i.e.

$$\Phi(x_{i,n}) = rx_{i,n}^p (1 - x_{i,n}^p). \qquad (17)$$

The dynamics of this map (Eq. (17)) are governed by two parameters, $p$ and $r$, which express intrinsic property of the environmental interfaces and the influence of the environment, respectively.

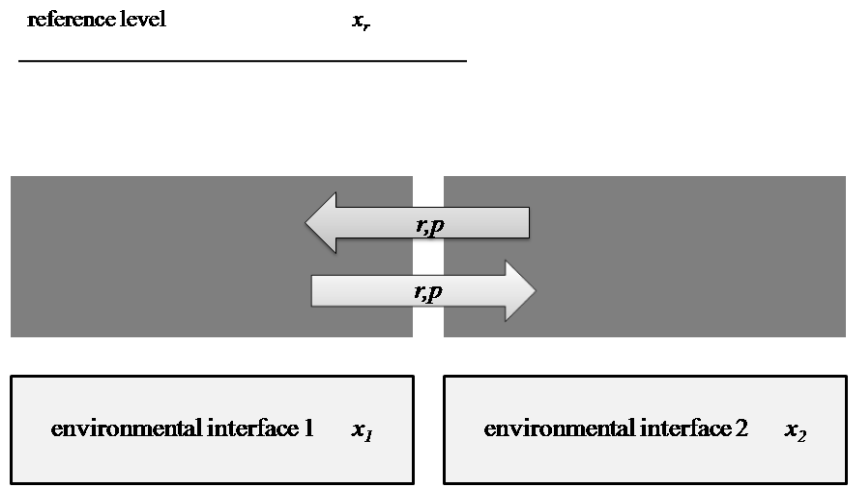

Figure 3. Schematic diagram of horizontal energy exchange between two environmental interfaces. Parameters $p$ and $r$ express intrinsic property of the environmental interfaces and the influence of the environment, respectively.

Since these and many other processes on environmental interface are defined as diffusion-like, we will explore (i) how these processes can be better represented in climate models by introducing parameter of exchange $p$ in the diffusive coupling associated with the horizontal energy exchange; and (ii) how the horizontal energy exchange intensity dynamics are affected by the perturbation of parameters that represent the influence of the environment, environmental interface coupling and horizontal energy exchange intensity.

In considering these problems we have to include observational heterarchy, a challenging topic when dealing with complex systems. Essentially, observational heterarchy reveals that it is impossible to unambiguously determine to which subsystems an element belongs [47]. Therefore, the dynamics of the complex system are articulated in terms of two kinds of dynamics, Intent and Extent dynamics, and the interaction between them, where Intent corresponds to an attribute of a given phenomenon and Extent corresponds to a collection of objects satisfying that phenomenon [47].

*3.2 Observational heterarchy and horizontal energy exchange between environmental interfaces*

Observational heterarchy consists of two sets of intra-layer maps, called Intent and Extent perspectives, and inter-layer operations satisfying the following conditions: (1) the inter-layer operations inherit the mixture of intra- and inter-layer operations and (2) there is a procedure by which the inter-layer operation can be regarded as an adjoint functor. If the inter-layer operation satisfies the conditions (1) - (2), it is called a pre-functor [47]. Preserving the above mentioned composition occurs as follows: A pre-functor, $\langle F \rangle$: Int $\to$ Ext is mapping a set, $X$, to a set, $\langle F \rangle X$, and map $\Phi$ to a map, $f^*\Phi f$, where $f^*f(x) = x$ for all $x$ in $f(X)$ with $f(X): \langle F \rangle X \to X$. In this sense we call $f^*$ pseudo-inverse of $f$. Because applying a pre-functor to a map is expressed as composition of maps, it satisfies the conditions (1) and (2). The approximation is defined by the assumption that $f$ is a one-to-one onto map. If one accepts the approximation, $f^* = f^{-1}$ holds, then a pre-functor can become a functor. Given two maps, $\Phi, \Psi : X \to X$

$$\langle F \rangle(\Phi)\langle F \rangle(\Psi) = (f^*\Phi f)(f^*\Psi f) = f^*\Phi(ff^*)\Psi f = \\ = f^*\Phi(ff^{-1})\Psi f = f^*\Phi\Psi f = \langle F \rangle(\Phi f). \quad (18)$$

It implies that $F$ preserves the composition of maps, $\Phi$ and $\Psi$.

The time development of the environmental surface dynamics $x_{i,n}$, for two interfaces, is expressed as

$$x_{i,n+1} = (1-c)\Phi(x_{i,n}) + f(\Phi(x_{j,n})), \quad (19)$$

where: $n$ is the time iteration, $i, j = 1, 2$, $x_{i,n} \in [0,1]$, $c$ the coupling parameter, $f$ the map representing the horizontal energy exchange between environmental interfaces, $\Phi$ is one of maps in the pair $(\Psi, \Phi)$ whose composition is preserved by a pre-functor $\langle F \rangle$. Here, we apply the framework of an observational heterarchy to the two environmental interface systems. If Intent and Extent are denoted by $\Phi$ and $\Psi$, respectively, the time development of the concentration is expressed as $x_{i,n+1} = (1-c)\Phi(x_{i,n}) + \Psi(x_{j,n})$. In this expression, if $\Psi(X) = f(\Phi(x))$ then it can be reduced to Eq. (19).

We perform our analysis following the procedure described in [47]. First, in this section we address the synchronization of the passive coupling for two environmental interfaces given by Eqs. (19) and (17), and then, in the next section, we will show that perturbation can modify the dynamics and enhance robust behavior in a multi-environmental interface system of active coupling. Synchronization is well-known collective phenomenon in various multi-component physical as well as the climate systems [48-50]. The exchange of information (coupling) among the components can be either global or local. Here, we consider that chaotic systems are synchronized only when the largest Lyapunov exponent of the driven system is negative. It was calculated by approach proposed in [51]. We studied the stability of the fixed point by linearzing $n \geq 2$ component coupled system, and obtain $\mathbf{Z}_{n+1} = \zeta_n \mathbf{Z}_n$ where $\zeta_n$ is the Jacobian of this system evaluated in $(0,0,\ldots,0)$ and $\zeta_n = (x_{1,n}, x_{2,n}, \ldots, x_{N,n})$. By iterating we obtain

$$Z_{n+1} = \left(\prod_{s=0}^{n} \zeta_s\right) Z_0, \qquad (20)$$

and thus we get Lyapunov exponent

$$\lambda = \lim_{n \to \infty} \left( \ln \left\| \prod_{s=0}^{n} \zeta_s \right\| \right) / n. \qquad (21)$$

Figure 4 depicts the diagram of normalized frequency of synchronization $F_p$ ($\lambda < 0$) for system of two environmental interfaces passively coupled (Eqs. (17) and (19)), as a function of parameter of exchange $p$, averaged over all values of the coupling parameter $c$ and logistic parameter $r$. The value of the normalized frequency of synchronization $F_p$ is calculated as

$$F_p = \frac{\sum N_n(\lambda < 0)}{\sum N_n(\lambda < 0) + \sum N_p(\lambda > 0)}, \qquad (22)$$

where $N_n(\lambda < 0)$ and $N_p(\lambda > 0)$ are numbers of negative and positive values of the Lyapunov exponent, respectively. These numbers were calculated for the fixed value of $p$, while $c$ and $r$ changing in interval $(0,1)$ and $(1,4)$ respectively, with the step of 0.05. From this figure it is seen that after $p > 0.2$, $F_p$ becomes lower, indicating a decrease of number of states, which are synchronized.

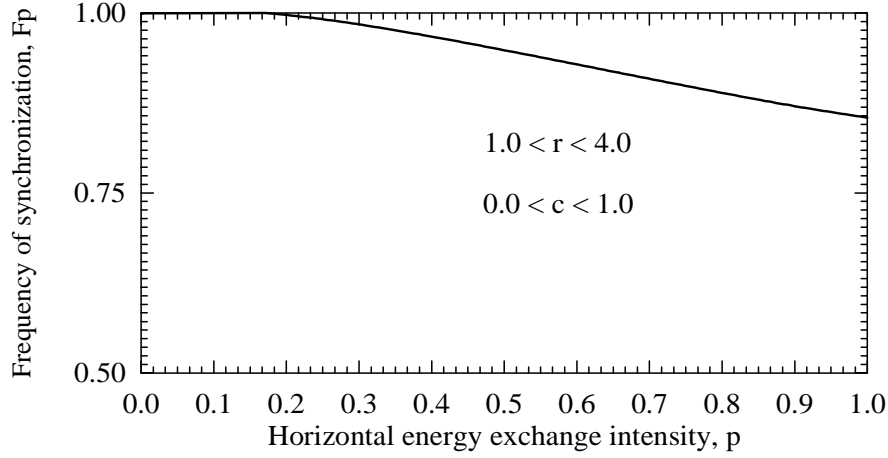

Figure 4. Normalized frequency of synchronization, $F_p$ ($\lambda < 0$) for system of environmental interfaces passively coupled (Eqs. (17) - (19)) as a function of parameter of exchange $p$. An averaging was done over all values of coupling parameter $c$ and logistic parameter $r$.

*3c Simulations of active coupling in a multi-environmental interface system*

Here, we address the behavior of active coupling [47], and estimate whether a coupled map system described above can achieve synchronization under influence of perturbations. The dynamics of two-environmental interface system called active coupling [47], used for simulations, is expressed as

$$x_{i,n+1} = (1-c)\Phi_n(x_{i,n}) + \Psi_n(x_{j,n}) \qquad (23a)$$

$$\Psi_{n+1} = f\Phi_n f^* f \qquad (23b)$$

$$\Phi_{n+1} = f^*\Psi_{n+1} \qquad (23c)$$

$$\Phi_n(x_{i,n}) = rx_{i,n}^p(1-x_{i,n}^p). \qquad (23d)$$

We note, that the dynamical system defined by Eqs. (23a) and (23d) is called the passive coupling, and that is a usual coupled map system. The active coupling can be approximated to passive coupling, where the approximation is defined by adjunction or the equivalence between Intent and Extent. Compared with passive coupling, the behavior of active coupling is much more complex [47]. In Eqs. (23a) - (23d), because of a pseudo-inverse map, $f^*$, all calculations are defined to be approximations. In simulations, the Intent map was a discontinuous map, expressed by $\Phi_{n+1} = f^*\Psi_{n+1}$.

In order to see how perturbation enhances robust behavior in the framework of observational heterarchy in a multi-environmental interface system represented by closed contour of coupled environmental interfaces exchanging the energy horizontally. Then the system of coupled difference equations for $N$ environmental interfaces exchanging the energy, can be written in the form of matrix equation

$$\mathbf{XN1} = (\mathbf{A} + \mathbf{B}) \cdot \mathbf{XN}. \tag{24a}$$

The elements in matrices in Eq. (24a) are

$$XN1_{i,n+1} = x_{i,n+1}, \quad XN_{i,n} = x_{i,n},$$
$$A_{i,k} = (1-c)\Phi_n(x_{i,n})\delta_{i,k} \tag{24b}$$

$$B_{i,k} = \begin{cases} \Psi_n(x_{k,n}) & k = i+1, i < N \\ 0 & k \neq i+1, i < N \\ \Psi_n(x_{k,n}) & k = 1, i = N \\ 0 & k \neq 1, i = N \end{cases} \tag{24c}$$

where $i = 1, 2, \ldots, N$ and $\delta_{i,k}$ is the Kronecker symbol.

Simulations with the active coupling, defined by Eqs. (23a)- (23d), were performed with and without perturbation given as in [47]. The results of simulations are shown in Figure 5. In this figure Lyapunov exponent $\lambda$ is plotted against coupling parameter $c$ for active coupling with perturbation (black line) compared to the passive coupling (red line), for different values of the parameter of exchange $p$ and the logistic parameter $r$. Simulations were performed with the closed contour of $N = 100$ interfaces. The Lyapunov exponent was calculated using Eqs. (20) - (21) and the Jacobian of the system given by Eqs. (24b)- (24c) is representing this contour.

In calculating $\lambda$, for each $c$ from 0.0 to 1.0 with step 0.005, $10^4$ iterations were applied for an initial state, and then first $10^3$ steps were abandoned. In order to see how the active coupling modifies the synchronization of horizontal energy exchange between environmental interfaces, we performed two kinds of simulations. Firstly, we used $r = 4.0$ and the fixed value of the parameter of exchange $p$ (Figures 5a - 5c); secondly, we used a randomly chosen $p$ and a logistic $r$ parameter with the values of 4.0, 3.82 and 3.6, respectively (Figures 5d - 5f). Figures 5a - 5c depict that in the chaotic regime ($r = 4.0$), regardless of the value $p$, the Lyapunov exponent is always positive ($\lambda > 0$) and therefore the process of the horizontal energy exchange in a multi-environmental interface system is always unsynchronized. However, the stormy perturbation disturbs this state (Figures 5a - 5c). Although the logistic parameter is settled at $r = 4.0$ for chaotic behavior, the coupling parameter $c$ tunes interaction and leads to synchronization in some intervals, particularly for $p = 1.0$ and $p = 0.5$. This behavior is more pronounced in Figures 5d - 5f where $p$ is randomly chosen; here the process of horizontal energy exchange in a multi-environmental interfaces exhibits a strong tendency towards the synchronization, even though the logistic parameter $r$ is in chaotic region ($r = 4.0$, 3.82 and 3.6).

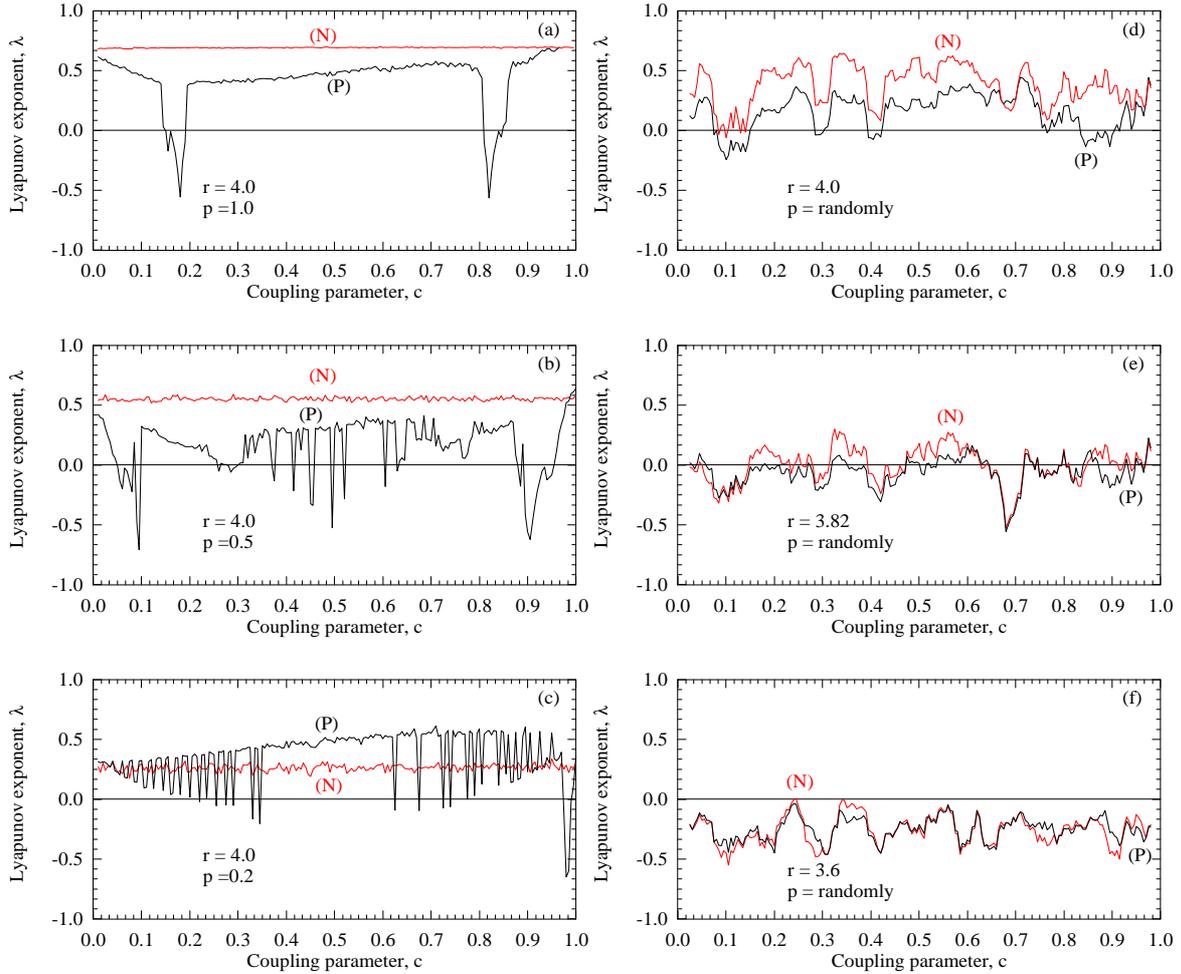

Figure 5. Diagram of Lyapunov exponent, $\lambda$, against coupling parameter $c$ for the fluctuated active coupling defined by Eqs. (23a - 23d) - P (black line) compared to passive coupling - N (red line) for different values of parameter of exchange $p$ and logistic parameter $r$. In (a)-(c) $p$ takes the fixed values (1.0, 0.5, 0.2), while $r = 4$. In (d)-(f) $p$ is randomly chosen, while $r$ takes values 4.0, 3.82, and 3.6 respectively. Simulations were performed with the closed contour of $N = 100$ environmental interfaces.

## 4. How to face the complexity of climate models

### 4.1 Background

In the introduction we have considered the complex ocean/atmosphere/land dynamical system, called weather and its long time average climate, as a complex one. This system is modeled by climate models having different levels of sophistication. The increasing complexity of those models is a growing concern in the modeling community. They are used to integrate and process knowledge from different parts of the system, and in doing so allow us to test system understanding and create hypotheses about how the system will respond to the virtual numerical experiments. However, if we strive to design our models to be more ''realistic'', we have to include more and more parameters and processes. Then, within this approach the model complexity increases, thus we are less able to manage and understand the model behavior. Obviously, the question about model complexity could be considered from the standpoint of a practitioner who sees it as a compromise between complexity and manageability. His\her question is basically very simple: ''How can I check if this model is appropriate to study this problem with this data set?'' According to Boshetti (2008): " As a result, the ability of a model to simulate complex dynamics is no more an absolute value in itself, rather a relative one: we need enough complexity to realistically model a process, but not so much that we ourselves cannot handle" [52].

Clearly, an answer to the above question requires: (i) a definition and a measure of complexity and (ii) that this measure is equally applicable to the model and to the data, because some sort of comparison is necessary. It is a hard task to find that measure even approximately. However, intuitively we can put a cadence on a view of complexity which is more related to a model's dynamical properties, rather than its architecture. Thus, we can say that in developing tools, which an advantage will be given to a tool which gives answers on the questions: (i) what is the maximal dynamical complexity a given model can generate? and (ii) what kind of different dynamical behaviors can a given model generate? as it is underlined by Boshetti (2008). For our consideration we will rely on Boshetti (2008) who defined the complexity of an ecological model as the statistical complexity of the output it produces that allows a direct comparison between data and model complexity [52]. Among the many different measures of complexity available in the literature, for that purpose, he adopted the statistical complexity defined in [53].

*4.2 An example of comparison between complexities of a global and regional model*

In this subsection we will illustrate an example of comparison between complexities of global and regional model. Here, we do not deal with statistical complexity of the global and regional models. Our intention is just to show possible differences in complexities of time series of precipitation as well as air temperature for both models, applying the algorithm for calculating the Kolmogorov complexity.

We have calculated the Kolmogorov complexity following Lempel and Ziv [54] who developed an algorithm for calculating the measure of complexity. It can be considered as a measure of the degree of disorder or irregularity in a time series. This algorithm performs the Kolmogorov complexity analysis of a time series $\{x_i\}$, $i = 1, 2, 3, 4, ..., N$ in the following way.

*Step 1:* Encode the time series by constructing a sequence $S$ of the characters 0 and 1 written as $\{s(i)\}$, $i=1,2,3,4,…, N$, according to the rule

$$s(i) = \begin{cases} 0 & x_i < x_* \\ 1 & x_i \geq x_* \end{cases} . \tag{25}$$

Here $x_*$ is a chosen threshold. We use the mean value of the time series to be the threshold. The mean value of the time series has often been used as the threshold [55]. Depending on the application, other encoding schemes are also used.

*Step 2:* Calculate the complexity counter $c(N)$. The $c(N)$ is defined as the minimum number of distinct patterns contained in a given character sequence. The complexity counter $c(N)$ is a function of the length of the sequence $N$. The value of $c(N)$ is approaching an ultimate value $b(N)$ as $N$ approaches infinity, i.e.

$$c(N) = O(b(N)), \quad b(N) = \frac{N}{\log_2 N}. \tag{26}$$

*Step 3*: Calculate the normalized complexity measure $C_k(N)$, which is defined as

$$C_k(N) = \frac{c(N)}{b(N)} = c(N) \frac{\log_2 N}{N}. \tag{27}$$

The $C_k(N)$ is a parameter to represent the information quantity contained in a time series, and it is to be a 0 for a periodic or regular time series and to be a 1 for a random time series, if $N$ is large enough. For a non-linear time series, $C_k(N)$ is to be between 0 and 1.

In order to calculate complexities of model time series we have used (i) air temperature and (ii) precipitation time series which are outputs from climate simulations for Belgrade and Novi Sad in Serbia [56, 57]. The Belgrade data set, for the period 2071-2100, was derived from: (a) the SINTEX-G which is a coupled atmosphere-ocean general circulation model [58] and (b) Eta-POM regional model [56]. The Novi Sad data set, for the period 2020-2050, was derived from: (a) the ECHAM5 which is the 5th generation of the ECHAM general circulation model [59] and (b) RegCM regional model [60].

We have calculated the Kolmogorov complexity for each time series obtained when each sample, in the original time series, is used as a threshold ($N = 10800$ for Belgrade and $N = 11323$ for Novi Sad). The results are depicted in Figure 6. We also have calculated Kolmogorov complexity (KL) and its maximal value (KLM) of time series from Figure 6. Results of those calculations are given in Table 1.

.

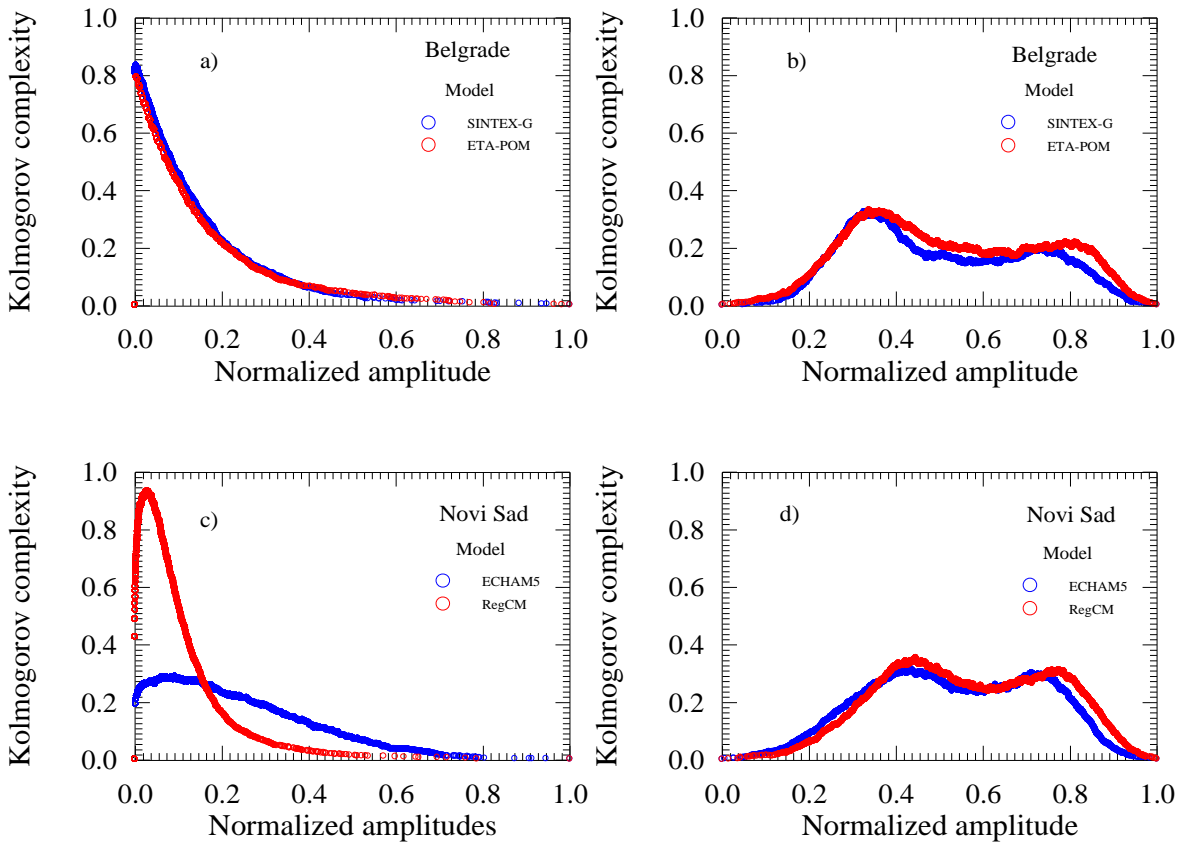

Figure 6. Kolmogorov complexity for the (a) precipitation and (b) air temperature time series for Belgrade and (c) precipitation and (d) temperature for Novi Sad, in Serbia, obtained from climate simulations using different models. On $x$ axis are depicted the values of the time series normalized as $x_i = (X_i - X_{min})/(X_{max} - X_{min})$, where $\{X_i\}$ is the time series of the precipitation or air temperature and $X_{max} = max\{X_i\}$ and $X_{min} = min\{X_i\}$.

From Figure 6a it is seen that there is no difference between complexities of the precipitation time series for Belgrade obtained by both models (global SINTEX-G and regional Eta-POM) over all amplitudes in time series. Moreover, the SINTEX-G model has slightly higher complexity. In contrast to that, Figure 6b depicts that the Eta-POM model mostly has the higher complexity than the SINTEX-G one for the air temperature time series. From Table 1 we can see that for air temperature time series the KL for the Eta-POM model (0.207) is higher than for the SINTEX-G model (0.176), while the KLM values are practically the same (0.331 and 0.326). Note that all of these complexities are pronouncedly low. Further inspection of this table clearly shows that the precipitation time series obtained by the SINTEX-G model, has higher complexities (KL - 0.705 and KLM - 0.834) than those obtained by the Eta-POM model (KL - 0.671 and KLM - 0.793). This analysis indicates the SINTEX-G and Eta-POM models, in particular for precipitation; have approximately the same level of complexity.

Now, we analyze the air temperature and precipitation time series for Novi Sad obtained by the global ECHAM5 and regional RegCM models. From Figure 6c it is seen that there is a large difference between complexities of the precipitation time series over all amplitudes in time series. Moreover, the RegCM model has pronouncedly higher complexity. Figure 6d depicts that the RegCM and ECHAM5 models mostly have very similar complexities for the air temperature time series. From Table 1 we can see that for air temperature time series the KL for the RegCM model (0.251) is higher than for the ECHAM5 model (0.241) and also for the KLM values - 0.354 and 0.318, respectively. Similarly, as for the above analyzed models, these values of complexity are still low. Further inspection of this table clearly shows that the precipitation time series obtained by the ECHAM5 model, has lower complexities (KL - 0.265 and KLM - 0.289) than those obtained by the RegCM model (KL - 0.871 and KLM - 0.935). This analysis indicates the ECHAM5 and RegCM models have approximately the same level of complexity in simulation of the air temperature. In contrast to that, there is a large difference in capabilities these models to simulate the participation.  To our knowledge this complexity analysis has not been used for analyzing the complexity of climate models.

|  |  | Model | | | |
|---|---|---|---|---|---|
|  |  | Global | | Regional | |
| Quantity | Measure | SINTEX-G | ECHAM5 | ETA-POM | RegCM |
| Temperature | KL | 0.176 |  | 0.207 |  |
| (Belgrade) | KLM | 0.326 |  | 0.331 |  |
| Temperature | KL |  | 0.241 |  | 0.251 |
| (Novi Sad) | KLM |  | 0.318 |  | 0.354 |
| Precipitation | KL | 0.705 |  | 0.671 |  |
| (Belgrade) | KLM | 0.834 |  | 0.793 |  |
| Precipitation | KL |  | 0.265 |  | 0.871 |
| (Novi Sad) | KLM |  | 0.289 |  | 0.935 |

Table 1. Kolmogorov complexities (KL and its maximum – KLM) values for the precipitation and air temperature time series for Belgrade and Novi Sad, in Serbia, obtained from climate simulations using different models.

## 5. Concluding remarks

We have considered some issues which are relevant for climate modeling. We gave a detailed overview of literature related to this subject. Then, we considered the climate modeling through the light of Gödel's Theorem that says that Number Theory is more *complex* than any of its formalization; further we clearly underlined the Rosen's definition of complexity and predictability. We have emphasized three issues.

Firstly, we have pointed out on occurrence of chaos in computing the environmental interface temperature from the energy balance equation when the given differential equation is replaced by a difference equation. For that purpose we have analyzed a coupled system of equations, often used in climate models. It is shown that the Lyapunov exponent mostly has positive values approving presence of chaos in this system, but there are still some strait regions where the Lyapunov exponent is negative and where the solutions of the coupled system are stable.

Secondly, we have analyzed the horizontal energy exchange between environmental interfaces which is described by the dynamics of driven coupled oscillators. To study their behavior and synchronization, when a perturbation is introduced in the system, as a function of the coupling parameter, the logistic parameter and the parameter of exchange, we have considered dynamics of two maps serving the diffusive coupling. Then, we have performed simulations, calculating the Lyapunov exponent, with the closed contour of $N = 100$ environmental interfaces.

Finally, we have explored possible differences in complexities of two global and two regional climate models using their output time series for the precipitation and air temperature. We have applied the algorithm for calculating the Kolmogorov complexity on those time series. We have found differences in the level of complexity among models.

**Acknowledgments**

This paper was realized as a part of the project "Studying climate change and its influence on the environment: impacts, adaptation and mitigation" (III43007) financed by the Ministry of Education and Science of the Republic of Serbia within the framework of integrated and interdisciplinary research for the period 2011-2014. The authors are grateful to the Provincial Secretariat for Science and Technological Development of Vojvodina for the support under the project "Climate projections for the Vojvodina region up to 2030 using a regional climate model" funded by the contract No. 114-451-2151/2011-01. The authors are grateful to professor D. Kapor for the invested effort and useful suggestions.